\documentclass{ECOC-author}
\usepackage[utf8]{inputenc}
\usepackage[T1]{fontenc}
\usepackage{graphicx}
\usepackage{xspace}
\usepackage[labelsep=space]{caption}
 
\usepackage[makeroom]{cancel}
\usepackage{amsmath,amssymb}

\usepackage{mathtools}
\let\floor\relax
\DeclarePairedDelimiter\floor{\lfloor}{\rfloor}

\usepackage{amsmath}
\usepackage{pgfplots}
\pgfplotsset{compat=newest}
\usepackage{tikz}
\usetikzlibrary{plotmarks}
\usetikzlibrary{positioning}

\pgfplotsset{every tick label/.append style={font=\tiny}}

\usetikzlibrary{calc}
\newlength{\MyFigureWidth}
\newlength{\MyFigureHeight}
\newcommand{\inputtikz}[1]{
      \includegraphics{#1.pdf}
      }
\pgfplotsset{
  every axis legend/.append style={
    legend cell align=left,
    align=left,
    font=\footnotesize
  }
}

\pgfplotsset{
  every axis plot/.append style={
      line width=1pt,
      mark size=1.5pt,
      mark options={solid,line width=0.5pt,fill=white!80!.}
    }
}

\pgfplotsset{
  every axis/.append style={
        scaled ticks = false, 
        tick label style={/pgf/number format/fixed},
    label style={font=\footnotesize},
        tick label style={font=\footnotesize}  
    }
}

\definecolor{UniformColor}{rgb}{0.00000,0.44700,0.74100}%
\definecolor{CapacityColor}{rgb}{0,0,0}%
\definecolor{CCDMInfColor}{rgb}{0,0.6,0}%
\definecolor{MPDMShortColor}{rgb}{0.85,0.325,0.098}%
\definecolor{MPDMMediumColor}{rgb}{0.929,0.694,0.125}%
\definecolor{MPDMLongColor}{rgb}{1,0,0}%

\pgfplotsset{Uniform/.style={dashed,color=UniformColor}}
\pgfplotsset{Capacity/.style={solid,line width=2pt,color=CapacityColor}}
\pgfplotsset{CCDMInf/.style={dotted,color=CCDMInfColor}}
\pgfplotsset{MPDMShort/.style={solid,color=MPDMShortColor, mark=diamond*}}
\pgfplotsset{MPDMMedium/.style={solid,color=MPDMMediumColor,mark=pentagon*}}
\pgfplotsset{MPDMLong/.style={solid,color=MPDMLongColor,mark=*}}
\definecolor{ColorConvMPDM}{rgb}{0.00000,0.44700,0.74100}%
\definecolor{ColorESS}{rgb}{0.4,0.8,0}%

\begin{document}


\title{\textsc{Analysis and Optimisation of Distribution Matching for the Nonlinear Fibre Channel}}%


\author{
    Tobias Fehenberger\corr and Alex Alvarado
}

\address{
   Department of Electrical Engineering, Eindhoven University of Technology, The Netherlands\\
   \email{tobias.fehenberger@ieee.org}
   }




  \begin{abstract}
    Enhanced Gaussian noise models are used to demonstrate that the per-block SNR after fibre transmission varies significantly due to the variable-composition nature of modern probabilistic shaping schemes. We propose a nonlinearity-optimised distribution matcher that improves the average and worst-case SNR by 0.14 and 0.22 dB, respectively.
  \end{abstract}
\maketitle                  

\section{Introduction}
\vspace*{-3pt}

  Probabilistic amplitude shaping (PAS) \cite{Boecherer2015TransComm_ProbShaping} has been evaluated in many different scenarios since its first demonstration in fibre optics \cite{Fehenberger2015OFC_ProbShaping}, and significant improvements over uniform quadrature amplitude modulation (QAM) have been found \cite{Buchali2016JLT_ProbShapingExp,Fehenberger2016PTL_MismatchedShaping}. The distribution matcher (DM) is a key building block of shaped communication systems based on PAS and has been studied in depth in order to enable high-throughput operation \cite{Pikus2017CommLetters_BLDM,Boecherer2017Arxiv_PDM,Fehenberger2019Arxiv_PASR}. Advanced DM schemes have been proposed that operate more efficiently, and thus, achieve lower rate loss than constant-composition distribution matching (CCDM) \cite{Schulte2016TransIT_DistributionMatcher}. Among these schemes is multiset-partition distribution matching (MPDM) for which various compositions are combined such that the target PMF is achieved on average \cite{Fehenberger2019TCOM_MPDM}. In enumerative sphere shaping (ESS) \cite{gultekin_enumerative_2019,Amari2019}, all compositions up to a certain maximum energy are used, which corresponds to using all shells inside an $n$-dimensional sphere. Note that we refer to ESS as a DM although the PMF cannot be controlled directly.

  The impact of signalling with probabilistic shaping on the magnitude of the nonlinear interference (NLI) of the fibre-optic channel has been investigated in theory, simulations, and experiments. Enhanced Gaussian noise (EGN) models \cite{Dar2013OptExp_PropertiesNLIN,Carena2014OptExp_EGNmodel} show that the magnitude of the NLI---and thus the effective signal-to-noise ratio (SNR)---depends on the standardized fourth moment (also known as kurtosis) of the QAM symbols, which has been confirmed in split-step simulations in \cite{Fehenberger2016JLT_ShapingQAM} and in experiments in \cite{Galdino2016ECOC_Experimental}. In \cite{Renner2017JLT_ShortReachShaping}, the impact of various probability mass functions (PMFs) on mutual information and SNR is investigated experimentally. In \cite{sillekens2018simple}, a PMF that is tailored to the nonlinear fibre channel is devised. A nonlinear shaping gain is obtained in \cite{Cho2016ECOC_ShapingNonlinearTolerance} by using a distribution that is on average flatter than the reference Maxwell-Boltzmann (MB) PMF.

  A common aspect of the aforementioned papers is that nonlinear performance is studied for \emph{average} symbol distributions and not for each DM block individually. This is a realistic assumption for constant-composition schemes where each shaped sequence has the same distribution, and also when many DM blocks are combined into a single, long forward error correction (FEC) codeword as the FEC decoder does not see the individual blocks but their average empirical distribution. When advanced DM schemes such as MPDM or ESS are used with very short FEC (up to the limit of a one-to-one length correspondence between DM and FEC), the statistics of the FEC codewords differ significantly and might never (or only rarely) follow the average PMF. For a linear channel with additive noise, the decoding performance will be approximately independent of this because the SNR does not depend on the PMF of each FEC block. For the nonlinear fibre channel, however, we have a different effective SNR for each sequence, even when intra-block effects (i.e., the ordering of the symbols within a shaped sequence) are ignored, as they are herein. Therefore, the decoding performance after fibre transmission can differ significantly for each block.

  In this paper, we study the finite blocklength regime of shaping for the nonlinear fibre channel in terms of per-block SNR variation. The analysis is based on the EGN model and shows significant SNR variations for ESS and MPDM. To reduce the adverse effect of fibre NLI, we propose, to the best of our knowledge, the first DM scheme whose per-block statistics are optimised for the nonlinear fibre channel.

\section{Principles of Distribution Matching}\label{ssec:review_dm}
\vspace*{-3pt}
A key building block in the PAS framework is the DM transforming a block of $k$ uniformly distributed bits to $n$ shaped amplitudes. Each output sequence is described by its composition $C=[n_1,n_2,\dots,n_m]$ denoting the number of occurrences $n_i$ of the $i$\textsuperscript{th} shaped amplitude. The number of uniquely addressable output sequences is given by the multinomial coefficient $\textrm{MC}(C)$. The number of input bits is then determined as $k = \floor{\log_2(\textrm{MC}(C))}$. A shaped scheme has a one-dimensional transmission rate of $k/n$, and its difference to the entropy of the shaped amplitudes is called rate loss, given in bits per 1D amplitude symbol (b/sym). All finite-length DMs suffer from such rate loss and advanced DM schemes aim to reduce it for a given block length $n$. In the following, we briefly review several DM schemes which are studied numerically in Sec.~\ref{sec:num_analysis}.

\begin{figure}
  \begin{center}
  \inputtikz{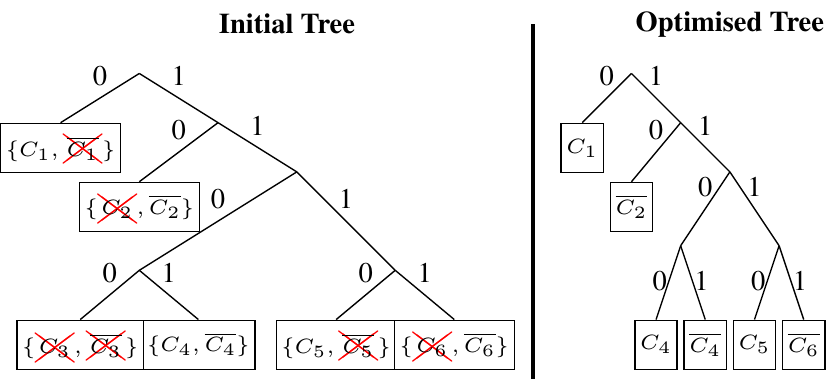}
  \end{center}
  \vspace*{-\baselineskip}
  \caption{Illustration of MPDM tree optimisation. The crossed-out compositions (left) have high NLI and are not used.}
  \vspace*{-0.5\baselineskip}
  \label{fig:tree_structure}
\end{figure}

For CCDM, each DM output block has the same composition $C_0$. Advanced schemes such as MPDM or ESS use many different compositions and thereby achieve a much smaller rate loss than CCDM. For pairwise MPDM \cite[Sec.~III]{Fehenberger2019TCOM_MPDM}, each composition $C$ has a complement $\overline{C}$ such that their combined average gives the target composition $C_0$. MPDM uses many of these pairs and arranges them in a binary tree such as the one shown in Fig.~\ref{fig:tree_structure}~(left). This construction achieves low rate loss while maintaining control of the average distribution. The transmission rate is determined only indirectly by the choice of distribution. ESS, in contrast, is operated at a certain rate and the shaped distribution is obtained indirectly by using all points inside a sphere, which gives an MB distribution as $n\to\infty$. While this is more energy-efficient than MPDM~\cite{gultekin_enumerative_2019}, it does not allow to control the average distribution or select the used compositions. We will show in Sec.~\ref{ssec:num_results} that this property can be disadvantageous for the nonlinear fibre channel.

\section{NLI-Optimised MPDM}\label{ssec:MPDM_optimisation}
\vspace*{-3pt}

In the following, we describe how MPDM can be optimised for the nonlinear fibre channel, which is, to the best of our knowledge, the first shaping scheme with NLI-optimised blockwise statistics. The idea behind the optimisation is that the EGN model \cite{Dar2013OptExp_PropertiesNLIN, Carena2014OptExp_EGNmodel} indicates that each DM block induces a different effective SNR after fibre transmission when the distribution of a block varies. This means that there are ``bad'' blocks that cause strong NLI and ``good'' blocks with low NLI.

For the proposed NLI-optimised MPDM, we refrain from using high-NLI compositions by setting their probability of occurrence to zero, i.e., by removing them from the binary tree shown in Fig.~\ref{fig:tree_structure}~(left). As this would reduce the transmission rate, we start the design process with MPDM that has one additional bit at its input, i.e., $k+1$ instead of $k$~bits. We then sort the compositions by the kurtosis of their PMF and remove the worst compositions (which have high kurtosis) until the number of inputs bits is reduced to $k$. This is illustrated in Fig.~\ref{fig:tree_structure} for a MPDM tree that is pruned to give an optimised MPDM where, in this example, the crossed-out compositions have high NLI and are eliminated. The pairwise structure of MPDM is broken up yet the general binary-tree structure is kept intact, see Fig.~\ref{fig:tree_structure} (right). Also, the average distribution is modified, which could for a linear channel without NLI, such as the additive white Gaussian noise channel, lead to a performance loss. We will show in Sec.~\ref{ssec:num_results} that for the nonlinear fibre channel, an improvement in both average and worst-case SNR is achieved by MPDM optimisation.

\section{Numerical Analysis}\label{sec:num_analysis}
\vspace*{-3pt}
\subsection{Fibre and Simulation Parameters}\label{ssec:params}
\vspace*{-5pt}

We use the EGN model to evaluate the effective SNR after long-haul fibre transmission. Four-wave mixing is neglected, and the number of Monte Carlo samples is set to 250000. The modulation format is dual-polarization 64QAM at 45~GBaud symbol rate. Root-raised cosine pulse shaping with 5\% roll-off is used. A wavelength-division multiplexing system with 11 channels spaced at 50~GHz is considered, and the centre channel is evaluated. The fibre link consists of 30 spans of 80~km standard single-mode fibre ($\gamma = 1.3$~1/W/km, $\beta_2 = -21.8$~ps\textsuperscript{2}/km, $\alpha=0.2$~dB/km) with an Erbium-doped fibre amplifier (5~dB noise figure) after each span. The considered figure of merit is effective SNR, which is computed according to the EGN formulas \cite[(16)--(18)]{Fehenberger2016JLT_ShapingQAM} for every composition.

\subsection{Assumptions for Numerical Comparison}\label{ssec:assumptions}
\begin{figure}[t]
\begin{center}
  \inputtikz{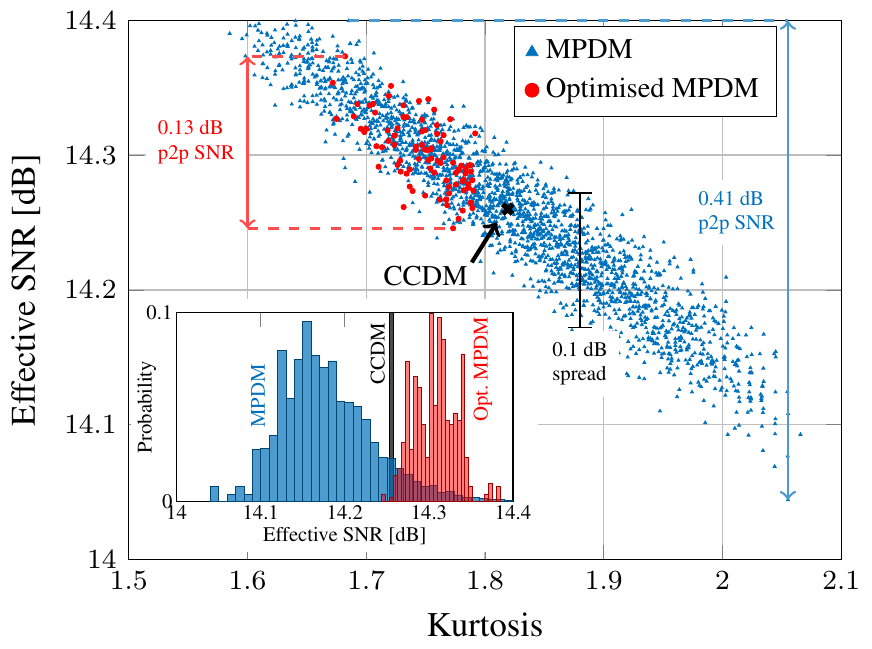}
  \vspace*{-\baselineskip}
  \caption{Effective SNR versus kurtosis for CCDM, MPDM and the NLI-optimised MPDM. Inset: Weighted histograms.}
  \vspace*{-\baselineskip}
  \label{fig:SNR_vs_KurtosisVec_30x80km}
\end{center}
\end{figure}

\vspace*{-5pt}
By using the EGN model to compare the per-block NLI of various shaping schemes in the finite blocklength regime, we make several assumptions that differ from the real fibre channel. Firstly, nonlinear interactions within each block are neglected. This means that the ordering of the shaped symbols within each DM output sequence is assumed to have no influence on the NLI. The impact of this assumption could be further studied with time-dependent fibre models such as \cite{Golani2016JLT_NLI}. We also note that the EGN model is derived for infinitely long blocks, while we use it for short blocks of a few hundred symbols. Also, adjacent blocks are assumed to not interact nonlinearly. Finally, the randomness in the numerical results is minimized as follows. Any variation in amplifier noise is neglected in the EGN analysis by computing the additive noise variance in closed form. The number of Monte Carlo samples for evaluating the EGN model is chosen large such that the reported SNR values can be attributed to fibre NLI with high confidence. As shown in Fig.~\ref{fig:SNR_vs_KurtosisVec_30x80km}, the SNR spread due to Monte Carlo simulations is approximately 0.1~dB for MPDM at a fixed kurtosis.

\subsection{Numerical Results}\label{ssec:num_results}
\vspace*{-5pt}

In the following, the effective SNR of various shaping schemes is investigated for the optical system outlined in Sec.~\ref{ssec:params}. The shaping block length is $n=216$ one-dimensional amplitude symbols and the number of input bits per block set to $k=349$, which gives a 2D transmission rate of $2\cdot k/n+1=4.23$~bits per QAM symbol. The block length is chosen to be compatible with the codes of the 802.11 standard of length 648~bits \cite{80211-2016}, resulting in a one-to-one relation between DM and FEC blocks. We note that the proposed methods work for any transmission rate supported by PAS.

In Fig.~\ref{fig:SNR_vs_KurtosisVec_30x80km}, effective SNR is plotted versus the kurtosis of CCDM and the MPDM schemes. CCDM, which uses in each block the same composition with kurtosis 1.82, has an effective SNR of 14.26~dB. We observe that MPDM has a much larger SNR variation than CCDM, with most blocks performing worse and some better than CCDM. Optimised MPDM (see Sec.~\ref{ssec:MPDM_optimisation}) has a smaller SNR fluctuation since compositions that cause low effective SNR are removed.
In the inset of Fig.~\ref{fig:SNR_vs_KurtosisVec_30x80km} the weighted histograms of the effective SNRs are shown, demonstrating clearly that both worst-case and average SNR are improved for optimised MPDM.

\begin{figure}[t]
\begin{center}
  \inputtikz{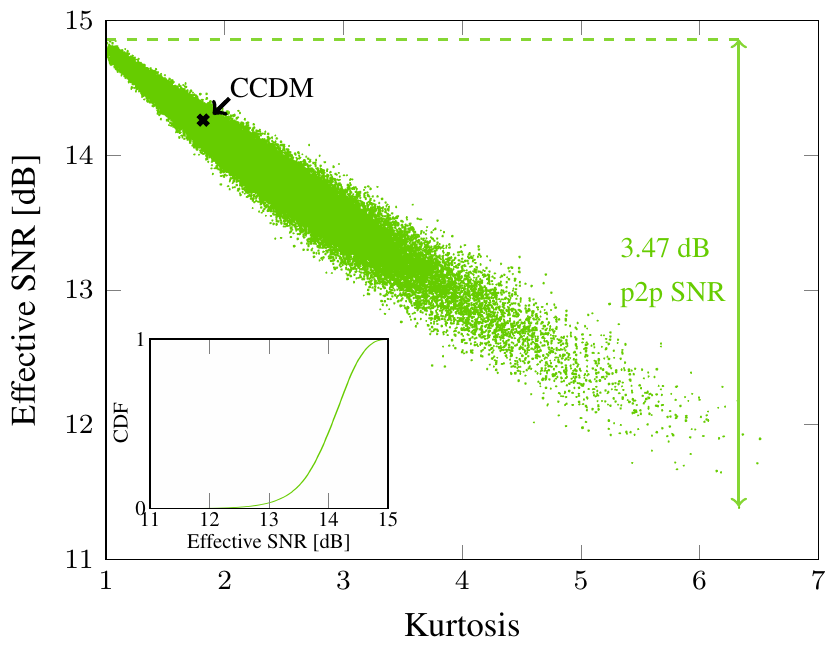}
  \vspace*{-0.2\baselineskip}
  \caption{Effective SNR versus kurtosis for enumerative sphere shaping. Inset: Cumulative distribution function (CDF).}
  \vspace*{-1.1\baselineskip}
  \label{fig:SNR_vs_KurtosisVec_ESS_30x80km}
\end{center}
\end{figure}

Figure~\ref{fig:SNR_vs_KurtosisVec_ESS_30x80km} shows effective SNR as a function of kurtosis for ESS. In comparison to MPDM, the increased variation in kurtosis, and thus the larger SNR fluctuation, are immediately observed. The reason for this is that ESS in general uses all compositions that are below a certain energy limit, including those that are highly unfavourable for the fibre channel. As of today, there is no known way of controlling the ESS construction to exclude these bad blocks. In the inset of Fig.~\ref{fig:SNR_vs_KurtosisVec_ESS_30x80km} the cumulative distribution function (CDF) of the SNRs is depicted, showing that only relatively few compositions lead to an SNR of less than 13~dB. Nonetheless, we argue that these low-SNR blocks can be a limiting factor to the overall system performance since they too must be decoded successfully.

In Table~\ref{table:comparison_30spans}, key numbers for a detailed comparison of the shaping schemes are given. CCDM is trivial since by always using the same composition, it achieves the same SNR under the assumptions stated in Sec.~\ref{ssec:assumptions}. MPDM has an average SNR of 14.17~dB, with a peak-to-peak (p2p) SNR variation of 0.41~dB (see vertical arrow in Fig.~\ref{fig:SNR_vs_KurtosisVec_30x80km}), and the worst-performing shaping block leads to an SNR of 14.03~dB. By optimising MPDM for the nonlinear fibre channel, the p2p SNR variation is significantly reduced to 0.13~dB SNR, and the average SNR is increased by 0.14~dB. Note that the worst-case SNR is improved by 0.22~dB, being now approximately identical to CCDM in SNR yet at a much lower rate loss of 0.021~b/sym instead of 0.053~b/sym. In addition to these improvements for the optical channel, we further note that the number of compositions is decreased for optimised MPDM, which could be beneficial in terms of storage requirements and to enable low-latency operation.
ESS by comparison uses a very large number of compositions and has a large SNR spread, as already discussed in Fig.~\ref{fig:SNR_vs_KurtosisVec_ESS_30x80km}. Although the best ESS blocks achieve an SNR that is the highest among all considered schemes and the ESS rate loss is the lowest, the worst-performing blocks have an SNR that is almost 3~dB worse than the low-SNR blocks of optimised MPDM. We argue that these unfavourably shaped blocks of ESS can have a significant impact on FEC performance. As they are the most likely to fail at decoding, they can be the limiting factor when a certain error rate is to be achieved.


\begin{table}
\caption{Nonlinear performance for various shaping schemes at\\ $n=216$. All SNRs, including peak-to-peak (p2p) SNR, are gi-\\ ven in dB. Rate loss is in bit/1D-symbol (b/sym).}
\begin{center}
\begin{tabular}{c|c|c|c|c}
 & CCDM & ESS & MPDM & Opt. MPDM \\
 \hline
\# compositions & 1 & 108066 & 2232 & 89 \\
 \hline
 rate loss & 0.053 & 0.014 & 0.02 & 0.021 \\
 \hline
max.~SNR& 14.26 & 14.86 &14.44 & 14.38 \\
min.~SNR& 14.26 & 11.39 &14.03 & 14.25 \\
 \hline
p2p SNR& 0 & 3.47 & 0.41 & 0.13\\
 \hline
average SNR& 14.26 &  14.24 &14.17 & 14.31\\
\end{tabular}
\label{table:comparison_30spans}
\end{center}
  \vspace*{-\baselineskip}
\end{table}

\section{Conclusion}
\vspace*{-3pt}
We have studied the finite-blocklength behaviour of various DM schemes in the presence of fibre nonlinearities. Modern DM schemes exhibit a significant SNR variation ranging from 0.4~dB (MPDM) to more than 3~dB (ESS). We believe that this NLI-induced fluctuation can have a significant impact on the FEC decoding margin since also the low-SNR blocks must be decoded successfully. We have further devised the first NLI-aware DM construction method. For the considered fibre setup, the optimised MPDM is found to give increases in average and worst-case SNR of up to 0.14~dB and 0.22~dB, respectively, at the same transmission rate as conventional MPDM.

\section*{Acknowledgements}
\vspace*{-10pt}
T.~Fehenberger would like to thank Y.~C.~G{\"u}ltekin (TU/e) for fruitful discussions on ESS. The work of A.~Alvarado has received funding from the European Research Council (ERC) under the European Union's Horizon 2020 research and innovation programme (grant agreement No 757791).

\section*{References}
\vspace*{-3pt}

\end{document}